\documentclass[%
 reprint,
 amsmath,amssymb,
 aps,
]{revtex4-2}

\usepackage{graphicx}% Include figure files
\usepackage{dcolumn}% Align table columns on decimal point
\usepackage{bm}% bold math
\usepackage{xcolor}
\begin{document}

\preprint{APS/123-QED}

\title{Proximity to an orbital order with charge disorder state in optimally-doped \textit{RE}\textsubscript{5/8}Ca\textsubscript{3/8}MnO\textsubscript{3} perovskites} % Force line breaks with \\

\author{Ben R. M. Tragheim$^{1}$}

\author{Clemens Ritter$^{2}$}

\author{Mark S. Senn$^{1, }$}%
 \email{m.senn@warwick.ac.uk}
\affiliation{%
 $^{1}$Department of Chemistry, University of Warwick, Gibbet Hill, Coventry, CV4 7AL, UK\\
 $^{2}$Institut Laue-Langevin, 71 Avenue des Martyrs, CS20156, 38042 Grenoble Cédex 9, France\\ 
}%

\date{\today}% It is always \today, today,

\begin{abstract}

The evolution of charge and orbital ordering phenomena in optimally-doped \textit{RE}\textsubscript{5/8}Ca\textsubscript{3/8}MnO\textsubscript{3} (RECMO, \textit{RE} $=$ rare-earth) manganite perovskites has been investigated through average structure synchrotron x-ray and neutron powder diffraction techniques. We demonstrate the intricate relationship between the \textit{B}O\textsubscript{6} octahedral rotation magnitude and lattice strain distortions acting in this series and how they tune macroscopic signatures describing ordering behavior. Through careful symmetry-motivated crystallographic analysis, we show that for the range of RECMO compositions which famously contain maxima in the colossal magnetoresistance (CMR) response, their lattice strain states are in close proximity to that associated with a novel orbital order:charge disordered state we have recently unveiled in the quadruple manganite perovskites Na\textsubscript{1-\textit{x}}Ca\textsubscript{\textit{x}}Mn\textsubscript{7}O\textsubscript{12}. We establish that this order is the primary state which competes with the ferromagnetic metallic state which ultimately leads to phase coexistence and the emergence of CMR. Our results lend themselves to aiding a further understanding of how particular chemical complexities can control charge and orbital ordering phenomena, and also the general properties of manganite perovskites and other related systems \textit{via} strain effects.

\end{abstract}

\maketitle

%\tableofcontents

\section{\label{sec:level1}Introduction }

\begin{figure}
\includegraphics{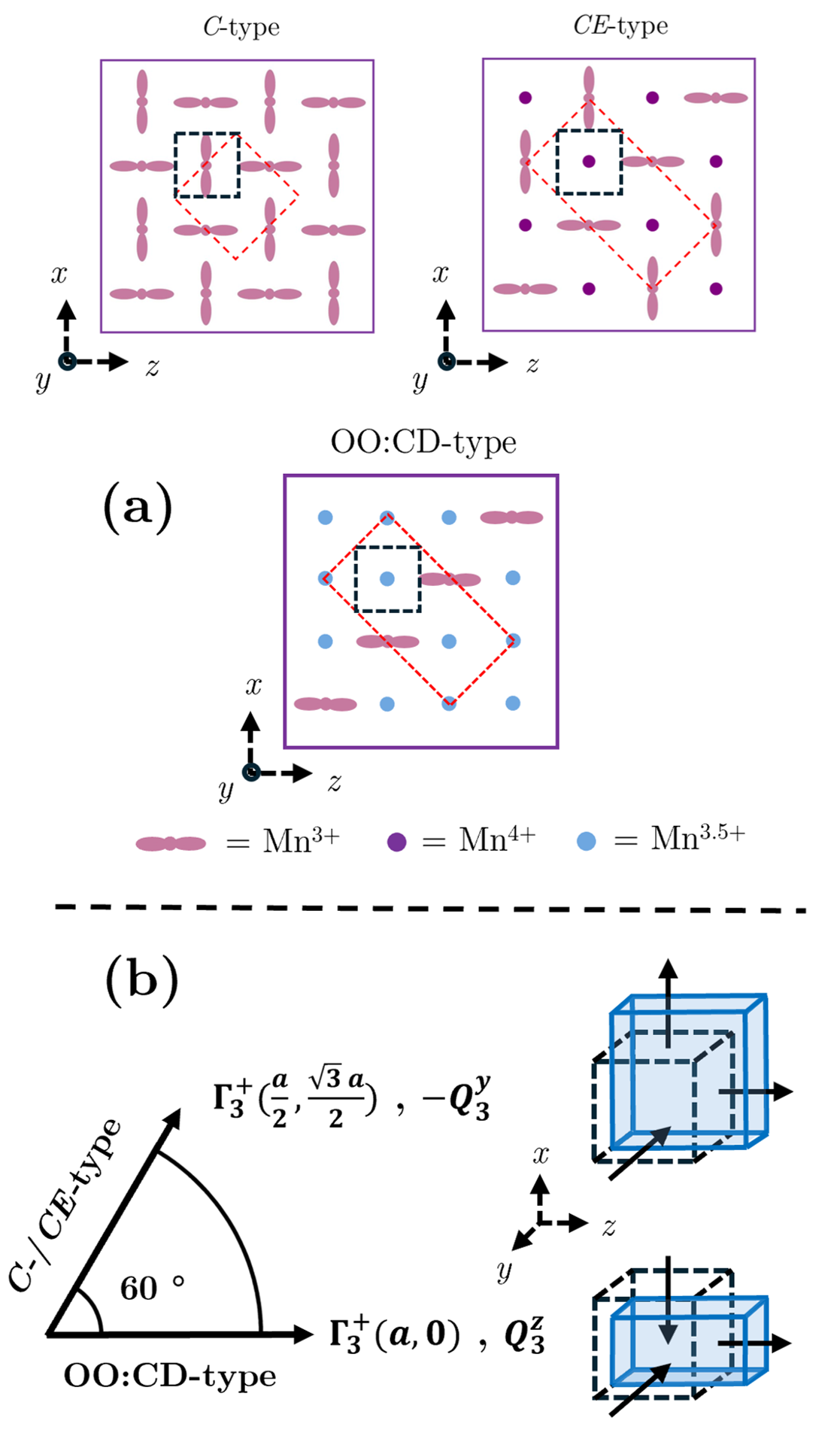}
\caption{\label{OO_scheme} (a) Illustration of the different types of charge and orbital order schemes observed in manganite perovskites. Each Mn site for each ordering scheme corresponds to the \textit{AB}O\textsubscript{3} perovskite \textit{B}-site, where \textit{A}- and O-sites are omitted for clarity. Red dashed boxes in each scheme represent the minimum repeating unit to describe ordering. Black dashed boxes indicate the unit cell of the \textit{Pm}$\bar{3}$\textit{m} \textit{AB}O$_3$ perovskite. All three schemes tile in the projection out of the plane. (b) The relationship between macroscopic strain transforming as the general irrep $\Gamma_3^+$, and its corresponding bases in order parameter space depicting the strain state of different charge and orbital ordering schemes. Black dashed lines represent the unstrained cell, and blue is the strained cell.} 
\end{figure}

Strongly correlated oxide systems, such as those found in cubic manganite perovskites (structural form \textit{A}MnO\textsubscript{3}), are hosts for observing intricate relationships between structural, electronic and magnetic degrees of freedom. Signatures for these are the emergence of charge and orbital order behavior, the periodic or local arrangement of cations (\textit{A}- or Mn-sites) of different oxidation states and \textit{d}$_\textit{z}^2$ orbitals of Mn$^{3+}$, respectively. The richness of charge and orbital ordering in the manganites is observed through the emergence of different unique ordering schemes, with examples including \textit{C}-type,\cite{wollan1955neutron, goodenough1955theory, bochu1974high} \textit{CE}-type\cite{goodenough1955theory, radaelli1997charge} and orbital order:charge disorder (OO:CD)-type,\cite{chen2021striping} shown schematically in Figure \ref{OO_scheme}(a). All of these ordering schemes have profound effects on the electronic and magnetic properties of their respective systems, and so understanding their presence and rationalising the microscopic mechanisms for their emergence allows for the successful tuning of their properties to aid in the design and optimisation of novel functional materials.  

One set of canonical manganite perovskites are those of the form La\textsubscript{1-\textit{x}}Ca\textsubscript{\textit{x}}MnO\textsubscript{3} (LCMO) and La\textsubscript{1-\textit{x-y}}Pr\textsubscript{\textit{y}}Ca\textsubscript{\textit{x}}MnO\textsubscript{3} (LPCMO) where for certain ranges of hole doping colossal magnetoresistance (CMR), the ability of a material to undergo large changes of resistivity under an applied magnetic field, is observed. A maximal CMR response occurs for precise doping and composition of \textit{x} $=$ \textit{y} $=$ 3/8,\cite{cheong2000colossal} which is believed to arise due to intrinsic macroscopic phase segregation between different insulating and metallic states, hosting respective charge and orbital ordered and disordered states.\cite{Uehara1} Hence, understanding the emergence of charge and orbital ordering phenomena for manganites at this doping level proves essential. However, adding to the existing complexity of the LCMO and LPCMO systems, continual hole-doping in these systems not only varies electronic states, but many other additional degrees of freedom are also unavoidably affected. Factors such as one-electron bandwidth narrowing,\cite{hwang1995lattice, fontcuberta1996colossal} \textit{A}-site cation size variance,\cite{rodriguez1996cation, vanitha1999effect} Jahn-Teller (JT) distortion tuning,\cite{VanAken1, beaud2009ultrafast, bovzin2007understanding} \textit{B}O\textsubscript{6} octahedral rotations,\cite{radaelli1997structural, barnabe1998role} phase segregation,\cite{Uehara1, kiryukhin2000multiphase, kim2000thermal} and strain\cite{ahn2004strain, baena2011effect} all contribute to preventing an accurate understanding for how any one particular control parameter affects the onset of charge and orbital ordering, and hence to optimize the CMR effect.    

To overcome the issues of these competing complexities, we have recently demonstrated the use of a prototype system for LCMO and LPCMO that aim to systematically tune only a single nominal control parameter, while keeping all others as invariant as possible. By using manganite quadruple perovskites Na$_{1-x}$Ca$_\textit{x}$Mn$_7$O$_{12}$ (structural formula \textit{AA'}$_3$\textit{B}$_4$O$_{12}$) we found that the state at \textit{x} $=$ 0.5, giving an average Mn \textit{B}-site oxidation state of $<$Mn\textsubscript{\textit{B}}$>$ $=$ $+$3.375 which is also coincident with the commonly observed optimal doping level of \textit{x} $=$ 3/8 in LCMO and LPCMO, corresponds to a novel  crystallographic state in which orbitally ordered stripes coexist with charge disordered layers (OO:CD-type scheme) \cite{chen2021striping}. The microscopic origin of this appears to be due to a coherent superposition between \textit{C}-type and \textit{CE}-type lattice modes leading to a cancellation of pronounced Jahn-Teller distortions in three out of every four layers.  This observation offers an explanation why the competition between metallic and insulating states, and hence CMR, is optimized at this point in the hole-doping phase diagram of the canonical manganites LCMO and LPCMO. Interestingly, due to the cubic structure adopted by the quadruple perovskites above the OO:CD-type state, the coupling between the electronic ordering and macroscopic strain was made particularly evident by our study of this prototype system. Our detailed analysis of the symmetry-adapted strain reveal a pronounced coupling to a tetragonal-type lattice elongation, contrasting strikingly with a tetragonal compression observed for doping levels that lead to \textit{C}-type and \textit{CE}-type orbital order (Figure \ref{OO_scheme}(b)). Our findings are thus indicative that the nature of the lattice strain states exhibited at this point in LCMO and LPCMO-type systems might well have significant interplay with the balance between the coexisting antiferromagnetic insulating and ferromagnetic metallic states, whose varying percolation under magnetic field ultimately gives rise to the observed CMR.

Previous models for various \textit{Pnma} \textit{AB}O\textsubscript{3} manganite perovskites with doping close to this optimal doping regime have been historically debated as to the nature and identity of charge and orbital ordering that occurs. It should be noted that the detailed structural investigations of LCMO and LPCMO around the \textit{x} $=$ 3/8 level are obfuscated by additional complexities such as microtwinning, electronic phase separation and high degree of pseudo symmetry, making extraction of reliable structural models difficult. Nevertheless, one model commonly adopted in the literature is the site-centred, \textit{CE}-type charge and orbital order scheme in which assignments were based on the observed ordering of half-doped La\textsubscript{0.5}Ca\textsubscript{0.5}MnO\textsubscript{3} below its charge order temperature, \textit{T}\textsubscript{CO} $=$ 250 K, consisting of a \textit{P}2\textsubscript{1}/\textit{m} space group where the average doping level differs by 1/8 compared to the optimal doping in LCMO and LPCMO.\cite{radaelli1997charge}.  This is also the same model found in our prototype system at \textit{x} = 0.5, \textit{A} = Na ($<$Mn\textsubscript{\textit{B}}$>$ $=$ $+$3.5). Another model, described in the literature is a bond-centred, `bi-striped' orbital ordering known as the Zener polaron, based on a crystallographic model derived from twinned single crystal data on Pr\textsubscript{0.6}Ca\textsubscript{0.4}MnO\textsubscript{3} below \textit{T}\textsubscript{CO} $=$ 235 K in \textit{Pm} space group.\cite{daoud2002zener} However, as of yet, no structural model for the $Pnma$-type perovskites consistent with the 3/8 doping level exists. Given the aforementioned pseudocubic lattice metric and complex microstructures associated with these compositions, complicating detailed structural investigation by high-resolution crystallographic techniques, it is maybe unsurprising that direct efforts to elucidate the nature of the charge and orbital ordering in the canonical LCMO and LPCMO manganites has been unsuccessful. Thus, highlighting the necessity for prototype systems to make progress in this area.   

The question then remains, to what extent do the prototype quadruple perovskites, specifically the novel OO:CD state identified at the 3/8 doping level, form a relevant prototype for LCMO and LPCMO. In this paper we hence target the system of \textit{RE}\textsubscript{5/8}Ca\textsubscript{3/8}MnO\textsubscript{3} (RECMO, \textit{RE} $=$ rare-earth) giving the doping level that coincides with that of LCMO and LPCMO producing maximal responses in the CMR effect.\cite{cheong2000colossal} Through the effect of varying the size of the \textit{RE}$^{3+}$ ionic radius has on the perovskite tolerance factor,\cite{shannon1976revised, goldschmidt1926gesetze} this prototype series provides the opportunity to study the interplay of various structural effects, such as \textit{B}O\textsubscript{6} octahedral rotations and lattice strains, that act to influence the observed electronic ground states at a fixed doping level. By careful Rietveld refinements and symmetry-mode analysis of structural models against synchrotron x-ray and neutron powder diffraction data, we show how the evolution of charge and orbital order in optimally-doped \textit{RE}\textsubscript{5/8}Ca\textsubscript{3/8}MnO\textsubscript{3} perovskites is systematically tuned through \textit{B}O\textsubscript{6} octahedral rotation distortions and their influence on the strain. While we find that the fitting of our diffraction intensities as a function of \textit{RE} composition is inconclusive concerning the exact nature of the structural distortions underpinning the orbital and charge order, the CMR response and signature for orbital ordering is clearly optimized at the point the symmetry adapted macrostrains most closely correspond to those observed in our prototype system within the OO:CD regime. Our analysis hence identifies this ordering as the one which is in competition with the ferromagnetic metallic state in the CMR perovskites.

\section{Experiment and data analysis}

RECMO samples (\textit{RE} $=$ La, La\textsubscript{4/8}Pr\textsubscript{1/8}, La\textsubscript{3/8}Pr\textsubscript{2/8}, La\textsubscript{2/8}Pr\textsubscript{3/8}, La\textsubscript{1/8}Pr\textsubscript{4/8}, Pr, Nd, Sm, Eu, Gd, Tb, Dy, Y, Er, Tm) were prepared by the solid-state synthesis method using a variety of conditions dependent on the identity of the rare-earth cation. Detailed synthesis conditions are described in the Supporting Information (SI). Characterization of samples \textit{via} synchrotron powder x-ray diffraction (PXRD) techniques was achieved using the high-resolution powder diffraction beamline I11 at the Diamond Light Source.\cite{Thompson1} Data were collected in the temperature range 100 K $\leq$ \textit{T} $\leq$ 300 K collected in approximately 2 K intervals using the position sensitive detectors (PSD). Data were also collected at \textit{T} $=$ 300 K and 100 K, above and below the charge order temperature (\textit{T}\textsubscript{CO}), respectively, using the multi-analyser crystal (MAC) detectors. A wavelength of $\lambda$ = 0.8240472(11) Å, as determined by refinement against an NIST Si standard, was used. Samples were packed into borosilicate capillaries and data collected in Debye-Scherrer geometry. Samples of RECMO (RE $=$ La, La\textsubscript{2/8}Pr\textsubscript{3/8}, Pr, Nd, Y) were characterized by powder neutron diffraction (PND) techniques using the high-resolution powder diffraction instrument D2B at the Institut Laue Langevin.\cite{hewat1986d2b} Data were collected at various temperature points between 10 K $\leq$ \textit{T} $\leq$ 300 K above and below \textit{T}\textsubscript{CO}. A wavelength of $\lambda$ = 1.59474(15) Å, determined from refinement against an NAC standard, was used. All diffraction data were analyzed \textit{via} the Rietveld refinement method with the software Topas Academic V7.\cite{Coelho1} 

Rietveld refinements were performed in the basis of irreducible representations (irreps) and symmetry-adapted formalism generated through the software ISODISTORT.\cite{Campbell1} Structural distortions of a lower symmetry structure can be classified as transforming as irreps of its parent structure (aristotype). In the following, the aristotype is chosen as \textit{AB}O$_3$ perovskite with a \textit{Pm$\bar{3}$m} space group and the \textit{A}-site placed at the origin. For compositions adopting the \textit{Pnma} perovskite structure, structural distortions transform as 8 different irreps. The full set of irreps and their order parameter directions (OPDs) are: R$_4^-$(\textit{a},\textit{a},0), R$_5^-$(-\textit{a},-\textit{a},0), X$_5^-$(\textit{a},-\textit{a};0,0;0,0), M$_2^+$(0;\textit{a};0) and M$_3^+$(0;\textit{a};0) which describe atomic displacements, and $\Gamma_1^+$(\textit{a}), $\Gamma_3^+$(\textit{a},-$\sqrt{3}\textit{a}$) and $\Gamma_5^+$(0,0,\textit{a}) which describe macroscopic strain. For compositions demonstrating charge and orbital order behavior, we have chosen to model structural distortions with respect to the site-centred charge and orbital ordered \textit{P}2\textsubscript{1}/\textit{m} structure with a basis [(1,0,1),(0,2,0),(-2,0,2)] and origin shift (0,1/2,0) with respect to the \textit{Pm}$\bar{3}$\textit{m} aristotype. This choice over the \textit{Pm} bond-centred structure will be made apparent in Section III. C. Most of the structural distortions of the \textit{P}2\textsubscript{1}/\textit{m} structure transform as the same irreps with respect to the \textit{Pm$\bar{3}$m} aristotype as observed in the \textit{Pnma} charge disordered structure. However, in order to account for the symmetry-lowering from orthorhombic to monoclinic symmetries, the extra strain irrep $\Gamma_3^+$(0,\textit{a}) and distortion mode $\Sigma_2$ become active. $\Sigma_2$ models the presence of superstructure peak formation with a propagation vector (1/4,1/4,0). For monoclinic splitting to occur, the ratio of $\sqrt{3}$ $\Gamma_3^+$(\textit{a},0) : $\Gamma_3^+$(0,\textit{a}) must be broken. Other distortion modes are allowed by symmetry in the \textit{P}2\textsubscript{1}/\textit{m} structure - namely $\Sigma_1$, S$_1$ and S$_2$ distortion modes - however, the effect of these are negligible and are considered to be inactive for RECMO that undergo charge order. For ease of comparison, we report all refined parameters, even those performed above the charge and orbital ordering phase transition temperature in $Pnma$, transformed according to the OPDs used in \textit{P}2\textsubscript{1}/\textit{m} setting. The full set of irreps and their OPDs used to generate \textit{P}2\textsubscript{1}/\textit{m} model are: R$_4^-$(\textit{a},\textit{b},0), R$_5^-$(\textit{a},\textit{b},0), X$_5^-$(\textit{a},\textit{b};0,0;0,0), M$_1^+$(0;\textit{a};0), M$_2^+$(0;\textit{a};0), M$_3^+$(0;\textit{a};0), M$_4^+$(0;\textit{a};0), $\Sigma_1$(0,0;0,0;\textit{a},0;0,0;0,0;0,0), $\Sigma_2$(0,0;0,0;\textit{a},0;0,0;0,0;0,0), S$_1$(0,\textit{a};0,0;0,0;0,0;0,0;0,0) and S$_2$(0,\textit{a};0,0;0,0;0,0;0,0;0,0) describing atomic displacements. $\Gamma_1^+$(\textit{a}), $\Gamma_3^+$(\textit{a},0), $\Gamma_3^+$(0,\textit{a}) and $\Gamma_5^+$(0,0,\textit{a}) describing lattice strain. 

Amplitudes of distortion modes are given in terms of absolute A$_p$ values, referring to the aristotype-cell-normalized amplitude of a particular distortion. The irrep notation has the form \textit{Q}(\textit{K}$_{\textit{n}}^{+/-}$) where \textit{Q} and \textit{K}$_{\textit{n}}^{+/-}$ denote amplitude and irrep of the distortion, respectively. The latter provides information on the \textit{k}-point in the aristotype Brillouin Zone (\textit{K}), the enumeration number (\textit{n}), and retention/violation of inversion symmetry for the site at the origin of the unit cell (+/-). For more details on other various intricacies of symmetry-mode analysis, the reader is directed to the work of Senn and Bristowe.\cite{Senn1}        

\section{Results and discussion}

\subsection{\label{sec:level2}300 K Datasets: Above \textit{T}\textsubscript{CO}}

Rietveld refinements of structural models against synchrotron PXRD data obtained at 300 K (Figure S1) demonstrate that all \textit{RE}$_{5/8}$Ca$_{3/8}$MnO$_3$ (RECMO) compositions contain a high phase purity and adopt the high temperature charge disordered \textit{Pnma} structure. Simultaneous refinements against both synchrotron PXRD and PND (Figure S2) data for the compositions \textit{RE} $=$ La, La\textsubscript{2/8}Pr\textsubscript{3/8}, Pr, Nd and Y, as well as the smooth variation in lattice parameter and unit cell volume as a function of tolerance factor (Figure S3) indicate that all compositions are sufficiently close to the nominal Ca$^{2+}$ doping and oxygen stoichiometry.

\begin{table}[!b]
\caption{\label{tab:table I} List of RECMO compositions investigated with their respective ionic radii and tolerance factor. Ionic radii are given for \textit{RE}$^{3+}$ in a IX coordination environment as this is the highest common coordination environment each \textit{RE} adopts. Radii labelled $^1$ are calculated as a weighted average of ionic radii between the nominal ratio of La$^{3+}$ and Pr$^{3+}$ for each corresponding \textit{RE}. Tolerance factor, \textit{t}, is calculated using the tolerance factor equation\cite{goldschmidt1926gesetze} using weighted average ionic radii values for an \textit{A}-site IX coordination environment of both \textit{RE}$^{3+}$ and Ca$^{2+}$ (1.18 Å), weighted average ionic radii values for a \textit{B}-site VI environment of high-spin Mn$^{3+}$ (0.645 Å) and Mn$^{4+}$ (0.53 Å), and a VI environment of O$^{2-}$ (1.4 Å).\cite{shannon1976revised}}
\begin{ruledtabular}
\begin{tabular}{ccc}
\textrm{\textit{RE}}&
\textrm{Ionic Radius (Å)}&
\textrm{Tolerance Factor, \textit{t}}\\
\colrule
La & 1.216 & 0.91926\\
La\textsubscript{4/8}Pr\textsubscript{1/8} & 1.209 $^1$ & 0.91763\\
La\textsubscript{3/8}Pr\textsubscript{2/8} & 1.201 $^1$ & 0.91599\\
La\textsubscript{2/8}Pr\textsubscript{3/8} & 1.194 $^1$ & 0.91436\\
La\textsubscript{1/8}Pr\textsubscript{4/8} & 1.186 $^1$ & 0.91273\\
Pr & 1.179 & 0.91109\\
Nd & 1.163 & 0.90756\\
Sm & 1.132 & 0.90072\\
Eu & 1.120 & 0.89807\\
Gd & 1.107 & 0.89520\\
Tb & 1.095 & 0.89255\\
Dy & 1.083 & 0.88990\\
Y & 1.075 & 0.88813\\
Er & 1.062 & 0.88526\\
Tm & 1.052 & 0.88306\\
\end{tabular}
\end{ruledtabular}
\end{table}

\begin{figure}[!t]
\includegraphics{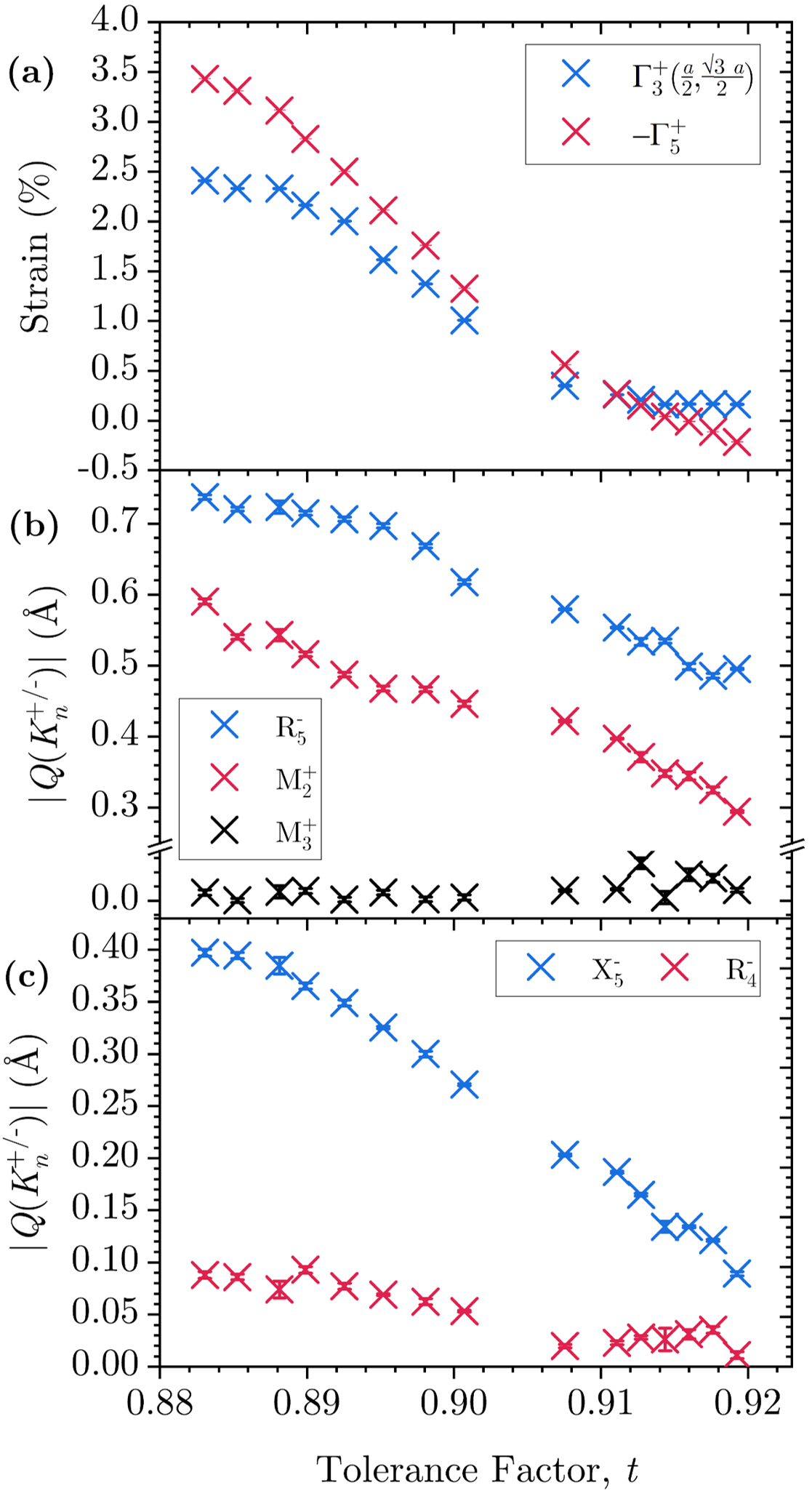}% Here is how to import EPS art
\caption{\label{300K-irreps} Variation of the structural distortions at \textit{T} $=$ 300 K that transform as irreps required to describe the charge disordered \textit{Pnma} structure for RECMO. (a) Distortions transforming as the irreps $\Gamma_3^+$($\frac{a}{2}$,$\frac{\sqrt{3}a}{2}$) and -$\Gamma_5^+$, given in $\%$. $\Gamma_5^+$ is given as its negative value in order to facilitate an easier comparison of strain evolution. (b) Structural distortions transforming as the irreps R$_5^-$, M$_2^+$ and M$_3^+$ describing the two unique \textit{B}O\textsubscript{6} octahedral rotation modes, and long-range \textit{C}-type Jahn-Teller distortion respectively, given in Å. (c) Structural distortions transforming as the irreps X$_5^-$ and R$_4^-$, given in Å. Points correspond to a single composition in RECMO, with the respective \textit{RE} identifier and tolerance factor value given in Table I. Errors of each mode are given for each data point, where some errors are smaller than the size of the data point.}
\end{figure}

Mode amplitudes for the \textit{Pnma} irreps extracted from Rietveld refinements of each RECMO member are shown in Figure \ref{300K-irreps} and are plotted as a function of tolerance factor, with the corresponding calculated tolerance factor values for each RECMO given in Table I. For a tolerance factor 0.8 $\leq$ \textit{t} $<$ 1, the ground state crystal structure of perovskites typically adopt an orthorhombic \textit{Pnma} GdFeO\textsubscript{3} structure, characterized by an \textit{a}$^-$\textit{b}$^+$\textit{a}$^-$ rotation of \textit{B}O\textsubscript{6} octahedra in Glazer tilt notation.\cite{glazer1972classification} In RECMO, the tolerance factor is varied \textit{via} decreases of the ionic radii of the rare-earth cation when considering \textit{RE}$^{3+}$ cations in a IX coordination environment across the lanthanide group,\cite{shannon1976revised} giving a physically tangible effect of increased magnitudes of \textit{B}O\textsubscript{6} octahedral rotations. The tuning of \textit{B}O\textsubscript{6} rotations through these tolerance factor effects is known to influence a number of different physical properties and phenomena, including magnetism described by superexchange interactions \textit{via} the Goodenough-Kanamori-Anderson rules.\cite{goodenough1955theory, kanamori1959superexchange, anderson1950antiferromagnetism}. As we will see, the tuning of these structural degrees of freedom \textit{via} the tolerance factor effect also has a strong influence of the magnitudes of symmetry breaking strains which themselves couple strongly the observed electronic ground states. 

\begin{figure*}[!t]
\includegraphics{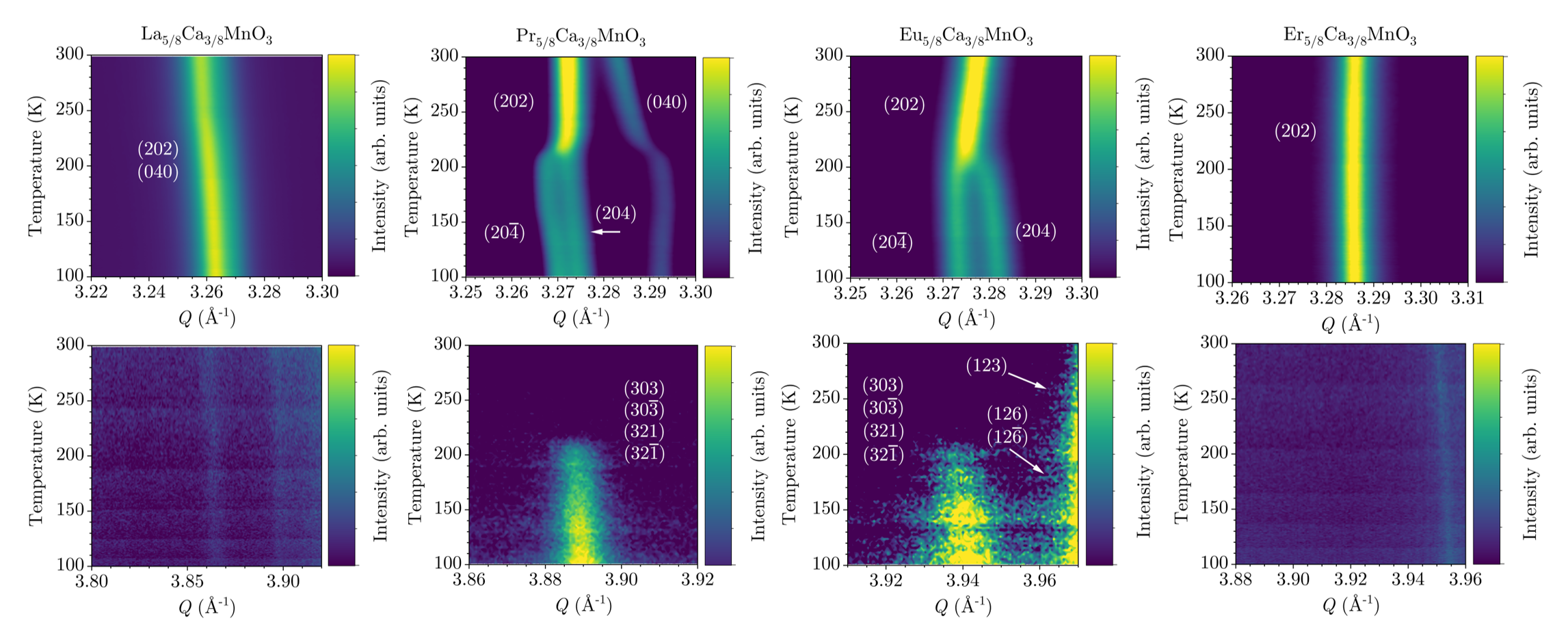}% Here is how to import EPS art
\caption{\label{RECMO-heatmaps} Heatmaps of variable temperature synchrotron PXRD data collected within the temperature range 100 K $\leq$ \textit{T} $\leq$ 300 K in approximately 2 K intervals. Data are plotted as a function of \textit{Q} ($\lambda$ $=$ 0.82405 Å) on the \textit{x}-axis to facilitate easier comparison with other RECMO compositions. Columns left to right correspond to RECMO compositions of \textit{RE} $=$ La, Pr, Eu and Er respectively. Top row of data indicate the evolution / absence of monoclinic splitting of the (202) \textit{Pnma} reflection to (20$\bar{4}$) and (204) \textit{P}2\textsubscript{1}/\textit{m} reflections. Bottom row of data indicate the evolution / absence of the superstructure peaks (303), (30$\bar{3}$), (321) and (32$\bar{1}$) in the \textit{P}2\textsubscript{1}/\textit{m} phase.}
\end{figure*}

The variation of tetragonal and shear strain, $\Gamma_3^+$($\frac{a}{2}$,$\frac{\sqrt{3}a}{2}$) and $\Gamma_5^+$ respectively, in Figure \ref{300K-irreps}(a) highlight the general observation that the magnitude of both strains increase with decreasing tolerance factor. The plateau-type behavior of both $\Gamma_3^+$ and $\Gamma_5^+$ close to 0 $\%$ within the tolerance factor window giving \textit{RE} $=$ La - Nd indicate that these compositions adopt, what has been referred to as, a pseudocubic (\textit{O}) state of the \textit{Pnma} structure over an orthorhombic (\textit{O'}) equivalent with a clearly evident lattice distortion. The importance of the \textit{O} state formation from an \textit{O'} state has been noted since it is at this point where various structural and physical phenomena begin to arise in manganite perovskites. Examples of this include the emergence of CMR in the LCMO solid-solution,\cite{Božin1, VanAken1} the high-temperature metal-to-insulator transition from orbital ordered to disordered states in LaMnO\textsubscript{3},\cite{Thygesen1, Carvajal1, Qiu1} and magnetic transitions \textit{via} Jahn-Teller distortion order-disorder behavior in LaMn\textsubscript{1-\textit{x}}Ga\textsubscript{\textit{x}}O\textsubscript{3}.\cite{Goodenough2, Blasco1, Farrell1} Figure \ref{300K-irreps}(b) highlights the amplitude evolution of the two unique octahedral rotation modes, R$_5^-$ and M$_2^+$, and the long-range \textit{C}-type Jahn-Teller distortion M$_3^+$. Here, the similar rate of increase for both R$_5^-$ and M$_2^+$ as a function of decreasing tolerance factor, and the invariant M$_3^+$ supports the fact that the RECMO system acts to tune the chemical complexity of \textit{B}O\textsubscript{6} rotations and isolates out effects due to Jahn-Teller distortions. The increases of X$_5^-$ and smaller gradient increases of R$_4^-$ antipolar \textit{A}-site displacements are shown in Figure \ref{300K-irreps}(c), highlighting their effect on the resulting ground state of the RECMO system. Combining these observations, we verify that RECMO acts as a suitable prototype system which systematically tunes the degree of \textit{B}O\textsubscript{6} octahedral rotation distortions and lattice strains that can be attributed to having a pronounced effect on the emergence of charge and orbital ordering. 
 
\subsection{\label{sec:level2}Variable Temperature Datasets: Probing \textit{T}\textsubscript{CO}}

In order to probe the presence of charge and orbital ordering phenomena in Ca$^{2+}$-doped \textit{A}MnO\textsubscript{3} perovskites two key macroscopic crystallographic signatures are required: monoclinic splitting and superstructure peak formation. These signatures occur due to the symmetry-lowering of the charge and orbital disordered orthorhombic \textit{Pnma} phase to a charge and orbital ordered monoclinic \textit{P}2\textsubscript{1}/\textit{m} phase, emerging at the charge ordering transition temperature \textit{T}\textsubscript{CO}. We perform variable temperature synchrotron PXRD measurements for each RECMO composition between 100 K $\leq$ \textit{T} $\leq$ 300 K to identify which compositions exhibit these signatures and their respective \textit{T}\textsubscript{CO}, demonstrated as heatmaps of data with examples for compositions of \textit{RE} $=$ La, Pr, Eu and Er shown in Figure \ref{RECMO-heatmaps}. All compositions are shown in Figure S4. 

In the case of \textit{RE} $=$ Pr and Eu, both signatures of monoclinic splitting and superstructure peak formation occur simultaneously in the temperature range 210 K $\leq$ \textit{T}\textsubscript{CO} $\leq$ 220 K, which in the case for Pr-based manganite perovskites is consistent with other compositions close to this optimal doping level.\cite{martin1999magnetic, lees1998neutron} Similar observations are noted for other optimally doped \textit{RE}\textsubscript{1-\textit{x}}Ca\textsubscript{\textit{x}}MnO\textsubscript{3} compositions for doping regimes 0.3 $\leq$ \textit{x} $\leq$ 0.4 of \textit{RE} that undergo charge ordering, all of which demonstrate a consistent charge order temperature range 190 K $\leq$ \textit{T}\textsubscript{CO} $\leq$ 230 K.\cite{hejtmanek1999interplay, richard1999room, kiryukhin2000multiphase} We also demonstrate in Figure \ref{RECMO-heatmaps} cases where both monoclinic splitting and superstructure peak formation signatures are not observed, as shown for \textit{RE} $=$ La and Er. These observations, coupled with our variable temperature PXRD data show in Figure S4, indicate that compositions of RECMO that do not undergo an orthorhombic to monoclinic transition within this temperature range are not likely to demonstrate charge and orbital ordering behavior below 100 K. Supporting this, PND data taken at 10 K for our compositions of RECMO for \textit{RE} $=$ La and Y (Figure S5), both of which are known not to undergo long-range charge and orbital ordering for doping regimes close to our optimal doping,\cite{dlouha2002structure, Uehara1} show the absence of any monoclinic splitting and superstructure peak formation due to crystal structure transformations. Hence, subsequent crystallographic analysis on charge and orbital ordering phenomena in RECMO will focus on data obtained at 100 K, sufficiently below the \textit{T}\textsubscript{CO} range observed in these optimally-doped manganite perovskites.        

\subsection{\label{sec:level2}100 K Datasets: Below \textit{T}\textsubscript{CO}}

Rietveld refinements of structural models against synchrotron PXRD and PND data obtained at 100 K are shown in Figure S6 and Figure S7, respectively. Each RECMO composition was modelled using the low symmetry \textit{P}2\textsubscript{1}/\textit{m} charge and orbital ordered structure, where appropriate symmetry constraints were applied to compositions that did not exhibit signatures of charge or orbital ordering. These symmetry constraints include fixing the amplitude of $\Sigma$$_2$ to 0 Å, and setting $\Gamma_3^+$(0,\textit{a}) $=$ $\sqrt{3}$ $\Gamma_3^+$(\textit{a},0). For compositions exhibiting charge and orbital order signatures, $\Sigma$$_2$, $\Gamma_3^+$(\textit{a},0) and $\Gamma_3^+$(0,\textit{a}) were allowed to refine independently. As noted in the introduction, compositions lying close to the optimal doping of various RECMO materials that undergo charge and orbital ordering have historically been modelled by either site-centred, \textit{P}2\textsubscript{1}/\textit{m} symmetry, \textit{CE}-type schemes or bond-centred, \textit{Pm} symmetry, ‘bi-striped’ schemes. While there might exist some contention into the determination of the exact scheme which is adopted,\cite{goff2004charge, rodriguez2005neutron, wu2007experimental, grenier2004resonant, efremov2004bond} and also that recent microscopy studies of Nd\textsubscript{0.5}Sr\textsubscript{0.5}MnO\textsubscript{3} thin films show in fact both schemes can coexist within the same sample,\cite{el2021charge} for the current study we have chosen to adopt the site-centred \textit{P}2\textsubscript{1}/\textit{m} scheme for modelling any ordering that occurs in RECMO. We choose this because the number of refinable atomic structural parameters contained within the \textit{P}2\textsubscript{1}/\textit{m} model is less than the \textit{Pm} model. Furthermore, in our previous work which attempts to fit the novel OO:CD-type ordering scheme to LPCMO compositions,\cite{chen2021striping} we refined structural models that are based on the site-centred \textit{P}2\textsubscript{1}/\textit{m} scheme, and for consistency we will use the same models here. A crucial point, however, is that the structural distortions in both site- and bond-centred models transform as the same irreps. This means that any evolution of the superstructure reflections will be well approximated regardless of the OPD used to generate the associated symmetry-adapted distortion modes. Mode amplitudes extracted from these Rietveld refinements are therefore given in Figure S8.

\begin{figure}[!t]
\includegraphics{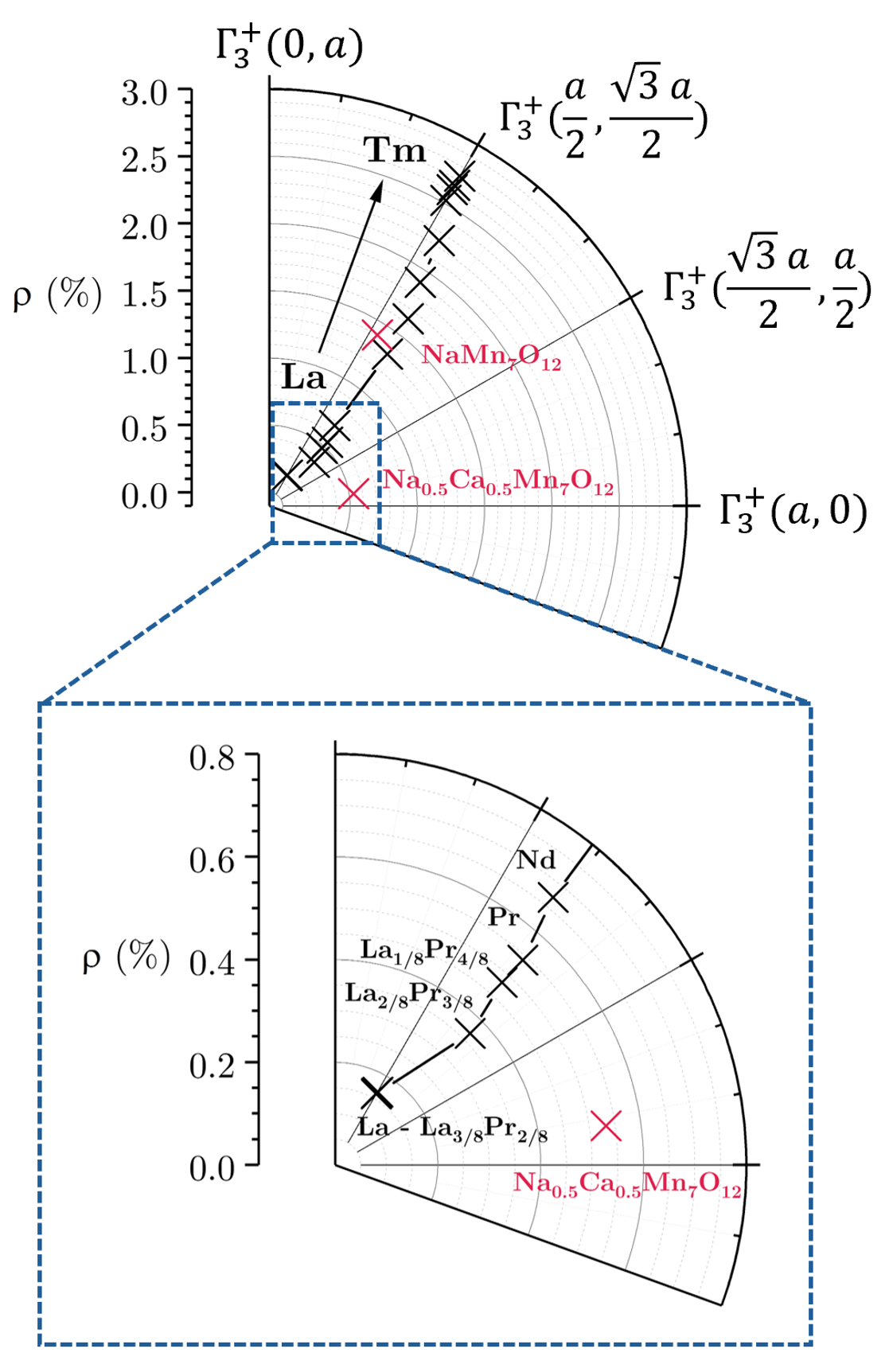}% Here is how to import EPS art
\caption{\label{strain_OPD} Polar plot of the order parameter space generated by $\Gamma_3^+$. Each point in the order parameter space corresponds to one composition in the RECMO system, with strain extracted from refinements of \textit{P}2\textsubscript{1}/\textit{m} models against diffraction data at 100 K. The angular ($\phi$) and radial ($\rho$) components for each point are calculated using formulae given in the text. Black crosses connected by black lines correspond to each of the 15 RECMO compositions with decreasing tolerance factor values from \textit{RE} $=$ La - Tm, demonstrated by the black arrow. Black crosses overlap for the compositions \textit{RE} $=$ La, La$_{4/8}$Pr$_{1/8}$ and La$_{3/8}$Pr$_{2/8}$. Red crosses correspond to data obtained from the strain variation observed in the OO:CD-type state found in Na\textsubscript{1-\textit{x}}Ca\textsubscript{\textit{x}}Mn\textsubscript{7}O\textsubscript{12} containing the same $<$Mn\textsubscript{\textit{B}}$>$ $=$ +3.375 as in RECMO and in NaMn\textsubscript{7}O\textsubscript{12} $<$Mn\textsubscript{\textit{B}}$>$ $=$ +3.5, corresponding with \textit{CE}-type ordering \cite{chen2021striping}. A zoomed-in region illustrating the proximity of RECMO compositions to the OO-CD-type strain state is given.}
\end{figure}

We show in Figure 4 how the variation of symmetry breaking strain $\Gamma_3^+$(\textit{a},0) and $\Gamma_3^+$(0,\textit{a}) across the RECMO series pertain to the breaking of orthorhombic to monoclinic symmetries within the basis of the \textit{Pnma} model (noting that with respect to the $Pm\bar{3}m$ perovskite these are not shear strains, see Figure \ref{OO_scheme}(b)). Figure \ref{strain_OPD} depicts the order parameter space of the general irrep $\Gamma_3^+$, where starting from the horizontal axis and rotating anticlockwise a new high-symmetry order parameter direction is reached every 30 $^\circ$. The horizontal axis corresponds to pure $\Gamma_3^+$(\textit{a},0) tetragonal strain (considered as a `1-long 2-short' distortion of the \textit{Pm}$\bar{3}$\textit{m} cubic perovskite lattice directions) and the vertical axis to pure $\Gamma_3^+$(0,\textit{a}) orthorhombic strain (a `1-long 1-medium 1-short' distortion of the cubic lattice directions). Note that the latter high-symmetry direction has no associated crystallographic constraint. This should not be confused with the orthorhombic symmetry observed in \textit{Pnma} perovskites which in fact arise from a combination of tetragonal $\Gamma_3^+$(\textit{a},0) and shear ($\Gamma_5^+$) strain, followed by a basis transformation.  Each point plotted in Figure \ref{strain_OPD} corresponds to one unique RECMO composition, where the angular ($\phi$) and radial ($\rho$) components are calculated by the equations: $\phi$ $=$ tan$^{-1}$[$\Gamma_3^+$(0,\textit{a})/$\Gamma_3^+$(\textit{a},0)] and $\rho$ $=$ $\sqrt{[\Gamma_3^+(\textit{a},0)]^2 + [\Gamma_3^+(0,\textit{a})]^2}$. There is a strikingly direct correspondence between the lattice distortions described using $\rho$ and $\phi$ from these strain modes to those describing the Jahn-Teller \textit{Q}\textsubscript{3} ('1 long 2 short') and \textit{Q}\textsubscript{2} ('1 long 1 medium 1 short') distortions of \textit{B}O\textsubscript{6} octahedral bond lengths derived from van Vleck modes.\cite{Vleck1, Kanamori1} Values of $\Gamma_3^+$(\textit{a},0) and $\Gamma_3^+$(0,\textit{a}) were obtained from Rietveld refinements of \textit{P}2\textsubscript{1}/\textit{m} structural models against powder diffraction datasets. Points that lie further away from the origin of the order parameter space indicate a greater magnitude of strain. 

Here, we note some key observations: firstly, compositions with a decreasing tolerance factor lie further away from the origin, consistent with an increasing magnitude of strain as observed in Figure \ref{300K-irreps}(a). Secondly, a subset of RECMO compositions with \textit{RE} $=$ La, La\textsubscript{4/8}P\textsubscript{1/8}, La\textsubscript{3/8}P\textsubscript{2/8}, Tb, Dy, Y, Er and Tm lie on the high symmetry axis $\Gamma_3^+$($\frac{\textit{a}}{2}$,$\frac{\sqrt{3}\textit{a}}{2}$). This high-symmetry axis corresponds to the case where $\sqrt{3}$ $\Gamma_3^+$(\textit{a},0) $=$ $\Gamma_3^+$(0,\textit{a}), the resulting crystallographic effect being that the monoclinic distortion angle is preserved as $\beta$ $=$ 90 $^\circ$, and hence diagnostic of the presence of the orthorhombic \textit{Pnma} charge disordered state. A subset range of RECMO compositions with \textit{RE} $=$ La\textsubscript{2/8}Pr\textsubscript{3/8} - Gd show to deviate from this high-symmetry order parameter space axis. In this instance, the crystallographic effect this has on the RECMO structure is to result in a monoclinic distortion angle $\beta$ $>$ 90 $^\circ$, which is indicative of the breaking of orthorhombic to monoclinic symmetries and hence the charge and orbital order transition emergence. Finally we note on the same diagram the position of our quadruple perovskite prototype systems at the nominal \textit{B}-site valence state $<$Mn\textsubscript{\textit{B}}$>$ $=$ +3.375 and $<$Mn\textsubscript{\textit{B}}$>$ $=$ +3.5 corresponding to OO:CD and \textit{CE}-type states. We see that, while for lower tolerance factors, the strain state remains nearest to that associated with the idealized \textit{CE}-type lattice distortion, at high tolerance factors, but slightly before the pseudocubic states at \textit{RE} $=$ La\textsubscript{1/8}Pr\textsubscript{4/8} and La\textsubscript{2/8}Pr\textsubscript{3/8}, a minimum distance to the idealized OO:CD strain state is reached. It is also interesting to note that, since only 60 $^\circ$ exist between the idealized strain state one would expect to be associated with OO:CD and that of a charge disordered \textit{Pnma} perovskite, there are three points in the order parameter space, $\Gamma_3^+$(0,0) (i.e. pseudocubic), $\Gamma_3^+$(\textit{a},0) and $\Gamma_3^+$($\frac{a}{2}$,$\frac{\sqrt{3}a}{2}$), (see Figure \ref{OO_scheme}(b)), that are equidistant. Consequently at a $\rho$ = $\frac{\sqrt{3}}{2}$ along the order parameter direction $\Gamma_3^+$($\frac{a}{2}$,$\frac{\sqrt{3}a}{2}$), the minimal distance with the point at $\Gamma_3^+$(\textit{a},0) is reached.  Considering the minimal distance between the \textit{Pnma} charge disordered state, and that associated with the idealized strain in OO:CD state ($\Gamma_3^+$(\textit{a},0) = 0.527 $\%$) we find that this lies at $\rho$ $=$ 0.456 Å which is closest to the RECMO compositions \textit{RE} $=$ La$_{1/8}$Pr$_{4/8}$ and Pr, even at 300 K, well above the charge and orbital ordering temperature.     

While the above discussion revolves around how the action of tetragonal and orthorhombic type-strains (in the basis of $Pm\bar{3}m$) compete, giving rise to monoclinic distortion (in the basis of $Pnma$), the shear strain ($\Gamma_5^+$) is equally important in capturing the departure of the RECMO series from pseudocubic symmetry as the tolerance factor decreases. To consolidate each of these three unique strains, we define the parameter of the Strain State (SS):
(SS=$\sqrt{[\Gamma_3^+(\textit{a},0)]^2 + [\Gamma_3^+(0,\textit{b})]^2 + [\Gamma_5^+]^2}$)
which describes the total strain that acts to distort the unit cell from the high-symmetry \textit{Pm}$\bar{3}$\textit{m} perovskite aristotype. Hence, this metric captures the deviation of perovskites from the pseudocubic state. Figure S9 demonstrates how SS varies across the RECMO series, indicating an increasing magnitude as a function of decreasing tolerance factor, similar to that shown in Figure \ref{300K-irreps}(a). Although SS tends towards zero for larger tolerance factors, the minimum (\textit{i.e.} compositions with closest proximity to the pseudocubic state) does not match either the maximum in $\Sigma_2$ (see Fig. S8), which is related to the signature for the \textit{CE}-type orbital order, or the maximum in the CMR response. However, a useful consequence of defining the SS is that this parameter can be used to directly compare the overall strain states of RECMO compositions to that of the quadruple perovskites Na\textsubscript{1-\textit{x}}Ca\textsubscript{\textit{x}}Mn\textsubscript{7}O\textsubscript{12} which are known to exhibit the OO:CD-type state,\cite{chen2021striping} and also because both unique systems contain the same strain irreps $\Gamma_3^+$(\textit{a},0), $\Gamma_3^+$(0,\textit{a}) and $\Gamma_5^+$. Figure S10 then demonstrates how SS varies across the Na\textsubscript{1-\textit{x}}Ca\textsubscript{\textit{x}}Mn\textsubscript{7}O\textsubscript{12} solid-solution, as a function of doping, with data being from the literature at \textit{T} $=$ 80 K, a temperature below \textit{T}\textsubscript{CO} for each composition. As noted in the introduction, quadruple perovskite compositions close to Na\textsubscript{0.5}Ca\textsubscript{0.5}Mn\textsubscript{7}O\textsubscript{12} which give a coincident $<$ Mn\textsubscript{\textit{B}} $>$ $=$ +3.375 as in RECMO give rise to the novel OO:CD-type state. Therefore, it is then pragmatic to compare the difference of SS for RECMO compositions and SS for Na\textsubscript{0.5}Ca\textsubscript{0.5}Mn\textsubscript{7}O\textsubscript{12} to act as an indicator for how close RECMO compositions are to the ideal strain state for observing OO:CD-type behavior.

\begin{figure}
\includegraphics{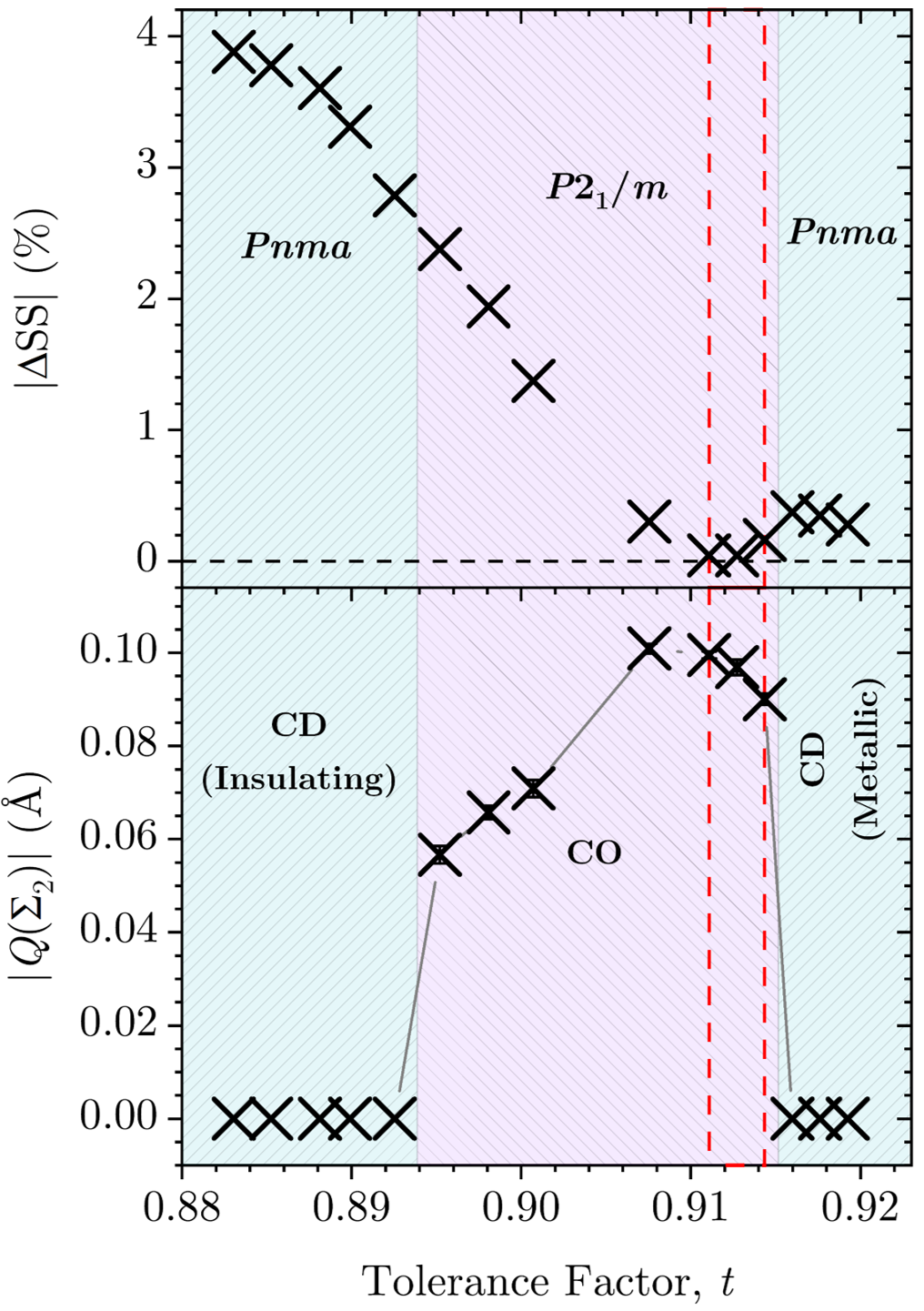}% Here is how to import EPS art
\caption{\label{delta-SS} (Upper) Evolution of the difference in Strain State, $|$$\Delta$SS$|$, between RECMO compositions at \textit{T} $=$ 100 K and Na\textsubscript{0.5}Ca\textsubscript{0.5}Mn\textsubscript{7}O\textsubscript{12} at \textit{T} $=$ 80 K\cite{chen2021striping} and (Lower) $\Sigma_2$ distortion mode amplitude for RECMO compositions at \textit{T} $=$ 100 K plotted as a function of tolerance factor. Both sub-figures contain approximate phase boundaries which depict the regions in which ordering transitions occur. The upper figure contains information on the change in space group symmetry between orthorhombic and monoclinic symmetries, and the lower figure contains information on whether RECMO compositions exhibit charge and orbital order or disordered states. CD $=$ charge-disorder, CO $=$ charge-order. A distinction is made in the CD states as being either metallic or insulating. The red dashed box indicates the range of compositions commonly associated to demonstrate a maximum in the CMR effect.\cite{cheong2000colossal, hwang1995lattice, hwang1995pressure}}
\end{figure}

The variation in the difference of SS for each RECMO composition at \textit{T} $=$ 100 K and Na\textsubscript{0.5}Ca\textsubscript{0.5}Mn\textsubscript{7}O\textsubscript{12} at \textit{T} $=$ 80 K, $|$$\Delta$SS$|$, as a function of decreasing tolerance factor is shown in Figure \ref{delta-SS}. Also shown in Figure \ref{delta-SS} is the evolution of the structural distortion $\Sigma_2$ (\textit{CE}-type mode) extracted from Rietveld refinements of \textit{P}2\textsubscript{1}/\textit{m} structural models against high-resolution diffraction data which directly models superstructure peak formation due to charge and orbital ordering behavior. A window which is commonly associated with compositions demonstrating a maximal in the CMR effect (\textit{RE} $=$ La\textsubscript{2/8}Pr\textsubscript{3/8} - Pr) is given by the red dashed boxes.\cite{cheong2000colossal, hwang1995lattice, hwang1995pressure} To the first approximation the occurrence of $\Sigma_2$ mirrors the evolution behavior of $|$$\Delta$SS$|$, as shown in Figure S11. As the RECMO series moves further away from the optimal strain state for compositions with lower tolerance factors $\Sigma_2$ is suppressed, but due to the increasing octahedral rotation distortions they remain insulating in character and adopt a charge and orbital disordered state. Effectively, in this low tolerance factor regime the underlying lattice strain appears to be frustrating any long range ordering. Similarly, at larger tolerance factors there is also a suppression of the signatures associated with long range charge and orbital order, however these changes are much more abrupt. The asymmetry of the $\Sigma_2$ `dome' likely reflects concomitant electronic band broadening which leads to rapid suppression of ordering signatures and the formation a metallic ground state. There is some evidence for this in Figure S8 where octahedral rotation modes R$_5^-$ and M$_2^+$ have a discontinuity across this region, that we suggest to be due to the coupling of the band structure, which is not observed at \textit{T} $=$ 300 K (see Figure \ref{300K-irreps}(b)). There are thus two competing order parameters at play: electronic bandwidth, which has been commented on previously,\cite{radaelli1997structural, medarde1995high} and the strain state we identify here as defined in relationship to the proximity to the ideal strain state associated with the OO:CD-type state observed in the quadruple perovskite equivalents.

The question then emerges: is this optimal strain state the causality or consequence of systems adopting the OO:CD-type state within a specific compositional window? Put another way, do we observe the smallest $|$$\Delta$SS$|$ because the lattice modes associated with the OO:CD-type state drive the system in this direction, or are the largest signatures of these modes observed at a given composition \textit{because} the $|$$\Delta$SS$|$ is already at a minimum prior to any electronic ordering taking place. To probe this we plot $|$$\Delta$SS$|$ in Figure S12 for strain extracted from refinements at \textit{T} $=$ 300 K, well above any electronic ordering. When comparing to the same plot but with $|$$\Delta$SS$|$ for strain extracted from refinements at \textit{T} $=$ 100 K (Figure 5), well below any electronic ordering, in general one can see that the compositions closest to $|$$\Delta$SS$|$ $=$ 0 $\%$ are the same above (300 K) and below (100 K) ordering temperatures. This therefore suggests causality in regions where the tolerance factor is not large enough to allow electronic band broadening effects to take over. However, all compositions within the red dashed lines in Figure 5 and Figure S12 - those in which CMR is generally considered to be optimized - move notably closer to $|$$\Delta$SS$|$ $=$ 0 $\%$ by \textit{T} $=$ 100 K, consistent with the strong coupling between the charge and orbital ordering and lattice strain observed in the quadruple perovskite equivalents. Identifying the nature of this coupling, we find by expanding the Landau free energy of the charge and orbital ordered state using INVARIANTS\cite{Hatch1} the fourth order coupling term  M$_3^+$($\Sigma_2$)$^2$[$\Gamma_3^+$(\textit{a},0) $+$ 1/$\sqrt{3}$ $\Gamma_3^+$(0,\textit{a})] proves to be a likely candidate in acting as an order parameter for controlling the competition between metallic and OO:CD-type ordering in RECMO. We may thus conclude that controlling the strain state of \textit{Pnma} perovskites observed at room temperature, compositionally, forms a valid strategy for enhancing the OO:CD-type state.  As is also evident from the discussion associated with Figure 4, $Pnma$ RECMO reaches the closest proximity to the OO:CD-type state, not for the most pseudocubic compositions, but for those with a small, yet clearly resolvable, tetragonal compression.

\section{Conclusion}

To conclude, using synchrotron PXRD and PND techniques coupled with symmetry-motivated crystallographic analyses we show that the onset of charge and orbital order behavior in optimally-doped \textit{RE}\textsubscript{5/8}Ca\textsubscript{3/8}MnO\textsubscript{3} (RECMO) is systematically tuned as a function of tolerance factor effects \textit{via} \textit{B}O\textsubscript{6} octahedral rotation distortions. We demonstrate the intricate relationship between the degree of octahedral rotation magnitude on the total strain state acting on RECMO compositions, and how this affects macroscopic signatures required to describe the formation of charge and orbital ordered states. By careful analysis, we have shown that the maximum in the CMR effect occurring in these optimally-doped manganites is coincident with a tolerance factor producing a strain state in closest proximity to that associated with the OO:CD-type state we have recently identified in quadruple manganite perovskites, demonstrating a pronounced '1 long 2 short' tetragonal lattice distortion. We thus establish this OO:CD-type state as the primary one competing with the ferromagnetic metallic state, which ultimately leads to phase coexistence and the aforementioned CMR at larger tolerance factors where electronic band broadening takes place. Crucially, this challenges conventional wisdom assuming \textit{CE}-type charge and orbital order which is inconsistent with the \textit{x} $=$ 3/8 doping level, and further assumed the effects of compositional tuning are realized in relationship to the proximity to the pseudocubic state. Our results will have a wide ranging impact for those seeking to control the properties of manganite perovskites and other related systems \textit{via} mechanical and epitaxial strain.

\begin{acknowledgments}
B. R. M. T acknowledges the University of Warwick and the EPSRC for studentship and funding (EP/R513374/1), and thanks Struan Simpson, Catriona Crawford and Matt Edwards for assistance in the collection of synchrotron x-ray and neutron diffraction data. M. S. S acknowledges the Royal Society for a fellowship (UF160265) and EPSRC grant "Novel Multiferroic Perovskites through Systematic Design" (EP/S027106/1) for funding. We are grateful to the STFC for the provision of synchrotron beam time at I11, Diamond Light Source, under the block allocation grant award (CY-32893), and to the ILL for neutron beam time at D2B under the proposal numbers 5-24-693 and 5-24-744 (DOI: 10.5291/ILL-DATA.5-24-693 and 10.5291/ILL-DATA.5-24-744). 
\end{acknowledgments}

% \nocite{*}

\bibliography{references}% Produces the bibliography via BibTeX.

\end{document}